\documentclass[runningheads,a4paper]{llncs}

\begin{document}

\title{Relations, Constraints and Abstractions:\\
Using the Tools of Logic Programming in the Security Industry}
\titlerunning{Logic Programming in the Security Industry}

\author{Andy King}

\institute{Portcullis Computer Security Limited, Pinner, HA5 2EX, UK}

\maketitle

\begin{abstract}
Logic programming is sometimes described as relational programming: a paradigm in which the programmer specifies and composes n-ary relations using systems of constraints. An advanced logic programming environment will provide tools that abstract these relations to transform, optimise, or even verify the correctness of a logic program.  This talk will show that these concepts, namely relations, constraints and abstractions, turn out to also be important in the reverse engineer process that underpins the discovery of bugs within the security industry.
\end{abstract}

\section{Introduction}

Logic programming is a wonderful paradigm: it is wonderfully expressive and yet also comes equipped with some wonderfully elegant semantics.  One legacy of the foundational work on semantics by pioneers such as Kowalski, Levi and van Emden, are suites of tools that we build and deploy within the field of research that we refer to as logic programming environments. Partial evaluators, program specialisation tools, and various program analyses are all formulated in terms of the base semantics proposed by these pioneers.  These base semantics provide a way to judge the correctness of a program manipulation technique, and by applying abstraction methods, we can even synthesise program analyses from these base semantics in a systematic and principled way \cite{cousot79systematic}.  Abstraction is a powerful idea in program manipulation, but when coupled with the pantheon of semantics that exist in logic programming, the concept becomes doubly powerful: we just need to select a suitably expressive semantics and then abstract it in an appropriate way. These ideas and these tools are so much a part of our heritage that we give this rich corpus of work a second thought.

The richness of the tooling that is available in logic programming becomes more evident when it is compared against the tooling that is available in reverse engineering. Reverse engineering is the discipline of extracting information from a program when the source is unavailable. Reversing engineering (abbreviated to reversing in the security sector) is routinely applied when performing a security audit on a commercial product that relies on software developed by a third-party such as a library. Security engineers also reverse to reason about the latest malicious programs and devise antivirus software. Reversing is also necessary when auditing programs for vulnerabilities that are introduced by the compilation process itself, or are most evident at the level of the executable. 

The most popular tool that is used for reversing within the security community is the IDA Pro dissembler \cite{pennell08reverse}.  This dissembler divides an executable into (more or less) its basic blocks, presenting them visually to the engineer in a flow diagram. Needless to say, the major impediment to reversing is the enormous effort required to understand an executable even when it is presented as a flow diagram.
As researchers in programming environments, we are conscious that tool support can underpin the development of a new program, and aid the understanding of an existing program. The problem of extracting information from a program --- which is the very essence of reversing --- is not new to us.  We recognise it as the problem of discovering invariants in a program. The problem is more about how to migrate techniques from higher-level paradigms to the level of an executable. In this short paper, we shall show how the familiar ideas from logic programming --- relations, constraints, and abstractions of constraints --- can be reinterpreted and reapplied in the setting of reverse engineering. 

\section{Where are the relations?}

The place to start has to be the base semantics. In assembler, the problem is not that the semantics is ambiguous (like some languages); the problem is more one of granularity. Instructions perform bit-wise operations on words rather than merely arithmetical and logical operations on variables.  The foci of computation are words, bit sequences and control-flags.  Moreover, these objects are referenced through pointers and pointer offsets rather than as local variables and as, say, elements of an array. These semantics can be modelled, at least partially, by using relations. The idea is to exploit the finite nature of machine words and model each word or register as a vector of bits. The before and after states of each instruction can then be represented as a relation between the bits of the input and output vectors. Such a relation can be described propositionally as a Boolean formula over the propositional variables in the input and output vectors. This approach of modelling is colloquially referred to as ``bit-blasting'' within the model-checking community, presumably because of the explosive nature of the technique.  Bit-blasting was famously used within the CBMC tool \cite{clarke04tool} in which the loops of C programs are unwound to a fixed depth so as to search for violations against prescribed correctness properties. Although initially treated with some skepticism, bit-blasting has gained acceptance as SAT solvers have emerged that can check the satisfiability of very large formulae  \cite{xie07saturn} and SMT solvers have been developed that include bit vector theories \cite{bryant07deciding} that directly support word-level instructions. 

Bounded model checking has been successfully applied to check invariants, and find circumstances in which invariants are violated, but it cannot extract a hitherto unknown invariant from a program. Nevertheless, the relational nature of bit-blasting does provide a base semantics that is compositional. To see this, consider a sequence of just two instructions that both add the constant one to the same 32-bit register. The input and output relation for this increment operation could be expressed as a Boolean formulae $f$ over the bits in the input and output vectors $\langle r_0, \ldots, r_{31} \rangle$ and  $\langle r'_0, \ldots, r'_{31} \rangle$ where $r_i$ and $r'_i$ are the variables that express the state of bit $i$ in the register before and after the increment. To compose two increments, two formulae $f_1$ and $f_2$ are obtained from $f$ by, respectively, systematically renaming the $r'_i$ variables to $r''_i$, and renaming the $r_i$ variables to $r''_i$.  The conjunction $f_1 \wedge f_2$ then asserts a double increment on the vectors $\langle r_0, \ldots, r_{31} \rangle$ and  $\langle r'_0, \ldots, r'_{31} \rangle$, albeit using a vector of temporary variables $\langle r''_0, \ldots, r''_{31} \rangle$. (The $\langle r''_0, \ldots, r''_{31} \rangle$ variables can be removed from the formula $f_1 \wedge f_2$ without loss of information by applying existential quantifier elimination.  This can give a denser representation of the composed semantics though it is not strictly necessary.)  By iterating this composition technique, it is possible to derive the relational semantics for a sequence of instructions of arbitrary length.

\section{Where are the constraints?}

One important idea in the analysis of logic programs is to use systems of constraints to describe systems of constraints \cite{giacobazzi95generalized}: systems of arbitrary Herbrand equations might be described by equations that are limited to depth-$k$; systems of finite domain constraints might be described by conjunctions of Horn formulae that express definiteness dependencies \cite{baker93definiteness}. We can reinterpret this idea for Boolean formulae and use formulae in one class to describe those in a more expressive class. Alternatively formulae could be described by systems of linear constraints. Using linear constraints as descriptions for formulae is more natural than one would initially think for reversing.  When formulae are derived by bit-blasting and composition, the relationship between input and output vectors often resemble systems of simple linear constraints. For instance, in the case of the double increment, the formula $f_1 \wedge f_2$ could be described by the constraint $2 + \sum_{i=0}^{31} {2^i}{r_i} = \sum_{i=0}^{31} {2^i}{r'_i} \bmod 2^{32}$. The constraint is not actually linear but is a congruent constraint with a modulo of $2^{32}$ which reflects the bounded nature of arithmetic that is expressed by the formula $f_1 \wedge f_2$. Describing the function with the linear relationship $2 + \sum_{i=0}^{31} {2^i}{r_i} = \sum_{i=0}^{31} {2^i}{r'_i}$ would actually misrepresent $f_1 \wedge f_2$. This is because, if the register initially stored the value $2^{32} - 1$, then after the double increment, the register will contain $1$ and not $2^{32} + 1$. Thus the relationship is only linear on a sub-range of the input data values. Congruence constraints are natural abstractions for reversing because they are already familiar to the reverse engineer.  This is because a number of security vulnerabilities relate to moduli; such vulnerabilities typically arise because the programmer has overlooked the wrap-around nature of arithmetic. Moreover, a security engineer will pay close attention to the size of an operand when reconstructing an algorithm from an executable.

An astute reader (and certainly a reverse engineer) will recall that a word can either be interpreted as a signed or an unsigned value.  The congruence $2 + \sum_{i=0}^{31} {2^i}{r_i} = \sum_{i=0}^{31} {2^i}{r'_i} \bmod 2^{32}$ stems from an unsigned treatment, otherwise congruence would be $2 + -2^{31}r_{31} + \sum_{i=0}^{30} {2^i}{r_i} = -2^{31}r'_{31} + \sum_{i=0}^{30} {2^i}{r'_i} \bmod 2^{32}$ where $r_{31}$ and $r'_{31}$ are the signs. However, observe that by adding $2^{32}r_{31} + 2^{32}r'_{31}$ to both sides, the congruence reduces to $2 + \sum_{i=0}^{31} {2^i}{r_i} = \sum_{i=0}^{31} {2^i}{r'_i} \bmod 2^{32}$. Thus the same congruence conveniently describes both the signed and unsigned interpretation of words.

Congruences reflect the bounded nature of computer arithmetic, but an equation such as $2 + \sum_{i=0}^{31} {2^i}{r_i} = \sum_{i=0}^{31} {2^i}{r'_i} \bmod 2^{32}$ possesses solutions for the variables $\langle r_0, \ldots, r_{31} \rangle$ and $\langle r'_0, \ldots, r'_{31} \rangle$ that are not 0-1 (truth) values. For instance, the congruence is satisfied by the assignment  $\{  r_0 \mapsto 2, r_1 \mapsto 0, \ldots, r_{31} \mapsto 0, r'_0 \mapsto 0, r'_1 \mapsto 2, r'_2 \mapsto 0, \ldots, r_{31} \mapsto 0 \}$. Such an assignment has no clear relationship with a Boolean function:  a Boolean function is characterised by its set of assignments to 0-1 values.  It is therefore necessary to be clear as to how a Boolean function can be described by a congruence. Formally, this is role of the concretisation map: the concretisation for system of congruences is the Boolean function whose satisfying assignments constitute the 0-1 solutions of the system (any solution that assigns a value other than 0 or 1 is simply ignored in this interpretation of a congruence).    

\section{Where are the abstractions?}

Stating the concretisation map (or dually an abstraction map) is much like providing a specification of a problem. Realising an algorithm that satisfies the specification and thus solves the problem is another thing entirely. Superficially it would seem that Boolean formulae and congruences are not closely related, and therefore it is not obvious how to find a system of congruences that best describe a given Boolean function. However, this problem can be recently solved using an iterative algorithm \cite{king08inferring}. The force of this result is that it gives a way to describe the relational semantics of an instruction, or even a sequence of instructions, with a system of congruences: bit-blasting is first used to derive a formula for the sequence and then this formula is described by congruences. Then the invariants on the basic blocks can be derived by fixpoint techniques \cite{muller07modular} that have been proposed for imperative programs.  To illustrate these ideas, we return to reasoning about a double increment.  For expositional purposes, we will suppose that a word is merely 4 bits wide. Then bit-blasting could derive the following system of (implicitly conjoined) formulae:

\[
f = \left\{ \begin{array}{l@{\qquad}l}
r'_0 \iff \neg r_0 &
r''_0 \iff \neg r'_0 \\
r'_1 \iff r_1 \oplus r_0 &
r''_1 \iff r'_1 \oplus r'_0 \\
r'_2 \iff r_2 \oplus (r_0 \wedge r_1) &
r''_2 \iff r'_2 \oplus (r'_0 \wedge r'_1) \\
r'_3 \iff r_3 \oplus (r_0 \wedge r_1 \wedge r_2) &
r''_3 \iff r'_3 \oplus (r'_0 \wedge r'_1 \wedge r'_2)
\end{array} \right.
\]

Note that the formula contains the intermediate variables $\langle r'_0, r'_1, r'_2, r'_3 \rangle$ which could be eliminated to derive a (possibly smaller) formula that still relates the input and output vectors $\langle r_0, r_1, r_2, r_3 \rangle$ and $\langle r''_0, r''_1, r''_2, r''_3 \rangle$.

A congruent description is derived for $f$ by first searching for a satisfying assignment (model) of $f$.  This can be readily accomplished with a SAT solver.  One such assignment is $M_1$ that is given below as 0-1 vector, where the propositional variables are ordered as follows $\langle r_0, r_1, r_2, r_3, r'_0, r'_1, r'_2, r'_3, r''_0, r''_1, r''_2, r''_3 \rangle$.

\[
\begin{array}{rcl}
M_1 & = & \langle 1, 0, 0, 0, 0, 1, 0, 0, 1, 1, 0, 0 \rangle \\
M_2 & = & \langle 1, 0, 0, 1, 0, 1, 0, 1, 1, 1, 0, 1 \rangle \\
M_3 & = & \langle 1, 0, 1, 0, 0, 1, 1, 0, 1, 1, 1, 0 \rangle \\
& \vdots & \\
M_{10} & = & \langle 1, 1, 1, 1, 0, 0, 0, 0, 1, 0, 0, 0 \rangle \\
\end{array}
\]

The truth assignment $M_1$ can be reinterpreted as the system of congruences $S_1$. For instance, the single assignment $r_0 \mapsto 1$ gives rise to the single congruence $r_0 = 1 \bmod 2^4$.  Henceforth, for brevity, we omit the modulo, which in this circumstance is chosen to be $2^4 = 16$ since words are 4 bits wide.

\[
\begin{array}{rcl}

S_1 & = &
\left \{ 
\begin{array}{l}
r_0 = 1, 
r_1 = 0, 
r_2 = 0, 
r_3 = 0, 
r'_0 = 0, 
r'_1 = 1, \\
r'_2 = 0, 
r'_3 = 0, 
r''_0 = 1, 
r''_1 = 1, 
r''_2 = 0, 
r''_3 = 0 
\end{array}
\right \} \\[2ex]

S_2 & = &
\left \{ 
\begin{array}{l}
r_0 = 1, 
r_1 = 0, 
r_2 = 0, 
r_3 = r'_3, 
r'_0 = 0, 
r'_1 = 1, \\
r'_2 = 0, 
r'_3 = r''_3, 
r''_0 = 1, 
r''_1 = 1, 
r''_2 = 0 
\end{array}
\right \} \\[2ex]

S_3 & = &
\left \{ 
\begin{array}{l}
r_0 = 1, 
r_1 = 0, 
r_2 = r'_2, 
r_3 = r'_3, 
r'_0 = 0, 
r'_1 = 1, \\
r'_2 = r''_2, 
r'_3 = r''_3, 
r''_0 = 1, 
r''_1 = 1 
\end{array}
\right \} \\[2ex]

& \vdots & \\[2ex]

S_{10} & = &
\left \{ 
\begin{array}{l}
r_0 + r'_0 = 1, 
r_1 + r''_1 = 1, 
4r_2 + 4r''_2 + 4 = 8r_3 + 4r'_0 + 4r'_1 + 8r'_2 + 8r''_3, \\
r'_0 + r''_0 = 1, 
2r'_1 + 4r'_2 + 2 = 8r'_3 + 2r''_0 + 2r''_1 + 4r''_2 + 8r''_3
\end{array}
\right \} \\

\end{array}
\]

The algorithm proceeds by searching for an assignment of $f$ that is not described by the system $S_1$.  This gives the model $M_2$ which can be translated into another system of simple congruences $S'_2$.  The system $S_2$ is then derived from $S_1$ and $S'_2$ by computing the merge of $S_1$ and $S'_2$. This is the unique system that contains all the solutions of $S_1$ and $S'_2$. This operation is not dissimilar to the affine hull that is used to merge systems of linear equations \cite{karr76affine}. With $S_2$ in place, the algorithm continues by searching for a model $M_3$ of $f$ that does not satisfy $S_2$.  Translating $M_3$ as a system of congruences gives $S'_3$ which is then merged with $S_2$ to give $S_3$ that is also given in the table. This iterative scheme continues until $S_{10}$ is derived.  All the models of $f$ are contained in $S_{10}$ and thus the algorithm stops at this point.

The system $S_{10}$ contains relational information pertaining to the intermediate bits as well as the input and output bits.  The intermediate bits can be eliminated by applying a triangular form \cite{karr76affine} which makes explicit any hidden relationships between the input and output bits:

\[
r_0 = r''_0, \quad
r_1 + r''_1 = 1, \quad
2r_1 + 4r_2 + 8r_3 + 2 = 2r''_1 + 4r''_2 + 8r''_3
\]

Interestingly, the relationships derived are richer than one would expect.  We have inferred that the states of the low bits are not changed by the double increment; that the states of the bits in position one always change; and that upper bits differ by two.

\section{Related work}

It has recently been pointed out that even recovering the control-flow graph is more complicated than one would initially expect \cite{kinder08jakstab} and, in fact, that IDA Pro often fails to reconstruct the complete control-flow graph. The problem stems in part from indirect calls, that is, when the address of a function is stored to a memory location pointed to by a register. The technical problem it is necessary to solve is to reason how intermediate instructions can possibly alter the value stored in the register and thereby infer that the address remained unchanged when the indirect call is resolved \cite{kinder08jakstab}. 

One notable body of work that also aims to support the reversing is the thesis work of Balakrishnan \cite{balakrishnan07wysinwyx}. Balakrishnan, under the direction of Reps, has developed a so-called value set analysis that attempts to uniformly track addresses and numeric values.  They intelligently chose a simple form of modulo constraint to represent a non-continuous range of values.  For example, in their notation 4[0, 12] denotes the set $\{ 0, 4, 8, 12 \}$ that describes the sets $\{ 0, 8 \}$ and $\{ 8, 12 \}$ among others. The rationale for this approach is that it enables sets of addresses on some word alignment to be accurately represented. We consider this approach to be a major advance in the analysis of binaries, since it attempts to seamlessly support addresses and numeric values. 

\subsubsection*{Acknowledgments}
This work was funded by EPSRC projects EP/C015517, EP/E033105 and EP/F012896
and a Royal Society Industrial Fellowship that has enabled King to be seconded to Portcullis Computer
Security Limited. We thank Harald S{\o}ndergaard who has contributed to much of this work.

\end{document}